\documentclass[11pt,twoside]{article}

\usepackage{aspstyle}
\usepackage{epsf}
\usepackage{lscape}

\def\rh{r_{\rm infl}}
\newcommand{\gap}{\;\rlap{\lower 2.5pt \hbox{$\sim$}}\raise 1.5pt\hbox{$>$}\;}
\newcommand{\lap}{\;\rlap{\lower 2.5pt \hbox{$\sim$}}\raise 1.5pt\hbox{$<$}\;}

\begin{document}
\title{Dynamical Models of the Galactic Center}   
\author{David Merritt}   
\affil{Rochester Institute of Technology}    

\begin{abstract} 
The distribution of late-type (old) stars in the inner parsec of the Milky
Way is very different than expected for a relaxed
population around a supermassive black hole.
Instead of a density cusp, there is a $\sim0.5$ pc core.
This article discusses what sorts of dynamical models might
explain this ``conundrum of old age.''
A straightforward interpretation is that the nucleus is
unrelaxed, and that the distribution of the old giants
reflects the distribution of fainter stars and stellar remnants
generally in the core.
On the other hand, a density cusp could be present in the 
unobserved populations,
and the deficit of bright giants could be a result of interactions with 
these objects.
At the present time, no model is clearly preferred.
\end{abstract}


The center of the Milky Way (MW) is special in terms of its location,
only 8 kpc away.
It is also home to perhaps the smallest supermassive black hole (SBH) with
a well-determined mass.
But in most respects, the center of our galaxy appears to be quite
ordinary when compared with the centers of other galaxies of 
comparable luminosity. 
It contains a dense nuclear star cluster (NSC) that
extends some ten parsecs from Sgr A$^*$ and that has a mass
of $\sim 10^7M_\odot$ \citep{SME:08}.
Population synthesis models suggest that star formation in the MW NSC
has been continuous over the last 10 Gyr, and sites of recent
star formation are apparent \citep{Figer:04a}.
These properties are typical of NSCs in other galaxies (B\"oker 2008).

Because of its proximity, the MW NSC can be resolved into individual
stars.
Number counts, together with reasonable guesses about the stellar mass
function, imply a density at 1 pc from Sgr A$^*$ of 
$\sim 10^5M_\odot \mathrm{pc}^{-3}$ \citep{Genzel:03,Schoedel:07}, 
and this density is consistent
with dynamical estimates based on stellar velocities
\citep{Oh:09}.
The implied, two-body relaxation time at 1 pc is roughly $10$ Gyr,
suggesting that there may have been enough time for the stars in the
inner parsec to have attained a relaxed, quasi-steady-state distribution
by now under the influence of random gravitational encounters. 
This assumption has been the basis for a great many
theoretical studies of the MW nucleus over the last two decades
(as summarized by T. Alexander in this volume).
In a relaxed nucleus, the distribution of stars and stellar remnants
is determined by just a handful of parameters: the total
density outside the relaxed region; the slope of the initial mass function; 
the mass of the SBH.

On the other hand, continuous star formation implies that at least
{\it some} stars in the MW NSC have been present for a time much less 
than the relaxation time.
This is clearly the case for stars in the two, parsec-scale 
stellar disks, which formed roughly 6 Myr ago \citep{Paumard:06,Bartko:09}.
Very recently, evidence has surfaced that even the old stars may not 
be relaxed.
Number counts of the late-type stars reveal a {\it core}, 
a region of essentially constant density near the SBH
\citep{BSE:09,Bartko:10}.
This is very different from the steep, power-law
density cusp expected in a relaxed nucleus \citep{BW:76}.
While it is possible that the observations are conspiring to 
mislead us -- there may still be a cusp in the fainter, 
unresolved stars, for instance --
the new data compel us to re-examine the assumption of a relaxed
steady state for the Galactic center.
Among the issues at stake is whether the MW
nucleus is well enough understood that it can serve as a 
template for other galaxies containing comparably-massive SBHs. 
These are the galaxies that would dominate the gravitational
wave signal 
as observed with space-based telescopes like LISA \citep{Hughes:03}.

\section{Some new (and not so new) puzzles}   

Recent observations reveal the following facts concerning
the inner parsec of the Milky Way.

\begin{itemize}
\item There is a core. Number counts of the late-type (old, cool) stars
show a well-defined inner break with respect to the $\Sigma\sim R^{-0.8}$
dependence at $R\gap 1$ pc
\citep{BSE:09,Bartko:10}.
Fitting of standard parametric models to the surface density
gives a core radius (the radius at which the surface density falls
to 1/2 of its central value) of $\sim 0.5$ pc (Fig.~1).
The core size is
independent of stellar luminosity down to the
current completeness limit of $m_K\approx 15.5$ mag,
corresponding to $1-3M_\odot$ red giants \citep{Dale:09}.
The deprojected (spatial) density profile $n(r)$ implies
$\lap 2000$ stars within $1$ pc of SgrA$^*$, although the form of 
$n(r)$ at $r\ll r_c$ is poorly constrained \citep{Do:09,Merritt:09}.

\item The distributed {\it mass} inside 1 pc is 
$1.0 \pm 0.5 \times 10^6M_\odot$.
This value is derived from proper motion velocities of a sample
of $\sim 6000$ stars in the projected inner parsec \citep{SME:09}.

\item Combined with the proper-motion mass estimate, measurement of
the diffuse light in the inner parsec implies a K-band mass-to-light 
ratio for the unresolved stars of 
$\sim 1.4_{-0.7}^{+1.4} M_\odot/L_{\odot, K}$ in this region
\citep{SME:09}.
This $M/L$ is consistent with an evolved stellar population having
a ``standard'' (Salpeter, Kroupa) IMF, in which a few percent
of the mass is in the form of stellar-mass black holes (BHs)
\citep{LBK:09}.
However given the uncertainties, it is also consistent with a 
somewhat larger remnant fraction.

\item Of the $\sim 200$ early-type (young, hot) stars in the
projected central parsec,
about half are Wolf Rayet and O stars occupying two stellar disks,
which appear to have formed in a well-defined event $6\pm 1$ Myr ago
\citep{Paumard:06,Lu:09,Bartko:09}.
The total mass associated with the disks is uncertain.

\item The luminosity function (LF) of these young disk stars shows a 
deficit at K magnitudes fainter than  $\sim 14$ mag,
compared with the K-band LF expected for a young population with a
standard IMF (Paumard et al. 2006; Bartko et al. 2010).
The ``missing'' stars are mostly main-sequence B stars.
One interpretation
is that the disk stars formed with a ``top-heavy'' IMF,
i.e. an IMF favoring massive stars.
The young stars in the central parsec that do {\it not} lie in the disks 
(the S-stars, and the young field stars) appear to
follow normal IMFs, with the expected predominance of main-sequence B-stars 
\citep{Bartko:10}.
\end{itemize}

\begin{figure}[!ht]
\plottwo{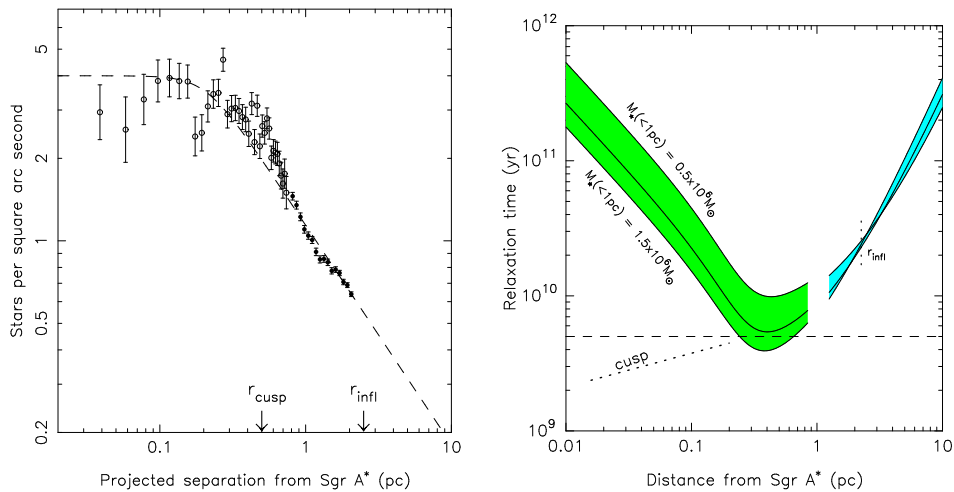}{merritt_fig1b.eps}
\caption{{\it Left:} Density of old stars at the Galactic center.
Open circles are binned counts of late-type  stars
brighter than $m_K=15.5$ mag \citep{BSE:09}.
Filled circles show the density of all stars with
$m_K\le 15.5$ mag and $R\ge 20''$ (Sch\"odel et al. 2007,
after corrections for crowding and completeness).
Dashed line is a broken-power-law model with 
$\Sigma\propto R^{-0.8}$ at large radii and inner slope of zero;
the core radius, defined as the radius at which the surface
density falls to 1/2 of its central value, is $0.49$ pc.
Arrows show the SBH influence radius and the expected outer radius
of the Bahcall-Wolf cusp.
{\it Right:} Estimates of the relaxation time, assuming a single-mass
population of Solar-mass stars.
Dashed horizontal line indicates the mean age of stars that formed
continuously over the last 10 Gyr.
Other details are given in the text.
}
\end{figure}

These observations are puzzling, and perhaps even inconsistent, 
for a number of reasons.

\begin{itemize}
\item There is no natural explanation for a parsec-scale core.
For instance, the radius at which red giants would be expected to
experience a collision with stellar-mass BHs, over their lifetimes,
is roughly an order of magnitude smaller than $r_c$,
even assuming that the BHs follow a steeply-rising,
relaxed density profile near the SBH \citep{Freitag:08}.
The assumption of a relaxed density profile in the BHs
is problematic however, given that...

\item There is no Bahcall-Wolf cusp in the stars.
If the late-type stars have been present for a time
longer than the two-body relaxation time $t_r$,
their distribution should  have relaxed by now to the
quasi-steady-state form $n\sim r^{-7/4}$ inside
$r_\mathrm{cusp}\approx 0.2\rh\approx 0.5$ pc, 
where $\rh\approx 2-3$ pc is the SBH influence radius
\citep{BW:76}.
The result would be a continuously-rising density of old
stars, not the essentially flat core that is observed.

\item The nuclear star cluster of the Milky Way,
on scales $1 \mathrm{pc} \lap r \lap 10$ pc,
appears to have undergone continuous star formation 
over the last 10 Gyr \citep{Mezger:99,Philipp:99,Figer:04b}.
If the top-heavy IMF inferred for the 
stellar disks is typical of past star formation in the core,
the mass-to-light ratio
in the inner parsec should be much higher than observed
by now, since a large fraction of stars would have evolved
to BHs (L\"ockmann et al. 2009).
Either star formation in the central parsec is just beginning,
which would make the current epoch special, or the IMF associated with
the event that formed the disks was atypical (or the inference
of a top-heavy IMF is incorrect; e.g. Bastian et al. 2010).
\end{itemize}

\section{Models}

Dynamical models of the inner parsec can be divided
into two broad classes.

1. {\sl Unrelaxed (low-density) models.} These models postulate 
that the low density observed in the late-type giants is 
characteristic of the old populations generally in the core,
including the fainter unresolved stars, and (possibly) the
stellar remnants.
In these models, the continued existence of a
core is consistent with 
the long relaxation time implied by a low density (Fig.~1).
Physical collisions between stars would be rare.

2. {\sl Relaxed (high-density) models.} A relaxed, 
Bahcall-Wolf cusp is assumed to be present, but for some reason 
it is not seen in the distribution of the red giants.
For instance, a high enough density of stellar BHs might
destroy the giants, or push them out from the center.

Low-density models suffer from a certain lack of robustness,
since it is easy to imagine mechanisms for refilling an empty core
(star formation, enhanced relaxation, etc.) and not so easy
to imagine ways of emptying it.
High-density models, on the other hand, 
are in danger of violating the proper-motion
constraint on the total mass in the core 
(by postulating too large a mass in BHs)
or the constraint on the mass-to-light ratio 
(by postulating too large a fraction of BHs relative to stars).

A key parameter in any model is the relaxation time,
which for a single stellar population is
\begin{eqnarray}
t_r &=& {0.33\sigma^3\over G^2 m\rho \ln\Lambda} \label{eq:tr}\\
&\approx& 1.5\times 10^{10}{\rm yr} 
\left(\frac{\sigma}{100\ \mathrm{km\ s}}\right)^3
\left(\frac{\rho}{10^5M_\odot{\rm pc}^{-3}}\right)^{-1}
\left(\frac{m}{M_\odot}\right)^{-1}
\left(\frac{\ln\Lambda}{15}\right)^{-1}; \nonumber
\end{eqnarray}
here $\sigma$ is the rms velocity in any direction,
$m$ is the mass of one star, 
$\rho$ is the mass density,
and $\ln\Lambda$ is the Coulomb logarithm.
If there is a range of mass groups, the concept of relaxation time
becomes vague, but a natural generalization is to
replace $m$ in equation~(\ref{eq:tr}) by $\tilde m$, where
$$
\tilde m \equiv {\int N(m) m^2 dm\over \int N(m) m dm} \nonumber
\label{eq:defmtilde}
$$
and $N(m)dm$ is the number of stars with masses in the range
$m$ to $m+dm$.
With this replacement, $t_r$ can be interpreted as the
time for a test star's velocity to be randomized by encounters with
more massive objects  \citep[e.g.][]{Merritt:04}.
Standard IMFs predict $\tilde m\lap 1M_\odot$;
if the density is dominated locally by stellar BHs,
$\tilde m\lap 10M_\odot$; if there is even a small population
of ``massive perturbers'' with $m\gg 10 M_\odot$, larger values of
$\tilde m$ are possible \citep{Perets:07}.

Ignoring for the moment the possibility of massive perturbers,
the relaxation time outside the core is quite well determined.
Fits to the stellar kinematics at $r\gap 1$ pc, together with the Jeans
equation, give a mass density
\begin{equation}
\rho(r) \approx \rho_0\left({r\over 1 {\rm pc}}\right)^{-1.8},\ \ 
 1 \mathrm{pc}  \lap r \lap 10 \mathrm{pc}
\label{eq:rhoi}
\end{equation}
\citep{Genzel:03,Schoedel:07,Oh:09},
with $\rho_0 \approx 1.5\times 10^5 M_\odot {\rm pc}^{-3}$; 
the uncertainty in $\rho_0$ is probably less than $50\%$. 
Figure 1 shows the implied $t_r$, assuming $m=1M_\odot$, 
for $\rho_0 = (0.75,1.5,3)\times 10^5 M_\odot {\rm pc}^{-3}$.
At the SBH influence radius, $r_\mathrm{infl}\approx 2.5$ pc,
the relaxation time is $\sim 2.5\times 10^{10}$ yr,
with a weak dependence on $\rho_0$.
Thus {\it assuming Solar-mass stars,
the two-body relaxation time at the influence radius of 
the Milky Way SBH is substantially longer than the age of the Galaxy},
and perhaps five times longer than the mean age of the stars.
This is neither a new, nor a controversial, result.
But it is worth emphasizing, since the time to establish a steady-state
Bahcall-Wolf cusp is approximately $t_r(r_\mathrm{infl})$
\citep{Preto:04,MS:06}.

The mass implied by equation~(\ref{eq:rhoi}) inside $1$ pc 
is $\sim 1.6\times 10^6M_\odot$ for
$\rho_0=1.5\times 10^5M_\odot$. 
This is somewhat larger than the
$\sim 1\times 10^6M_\odot$ inferred from the proper motions, 
but not so much larger that one can rule out the hypothesis that
the mass density continues to obey $\rho\sim r^{-1.8}$ inside the
observed core,
as it would if a Bahcall-Wolf cusp were present.

Inside $1$ pc, the relaxation time depends critically
on the assumed mass density and its variation with radius.
The latter is poorly constrained by the proper motion data 
\citep{SME:09}.
Figure~1 shows $t_r$ assuming that the mass is distributed 
as $\rho\sim r^{-0.5}$, the steepest dependence consistent with
the number counts of the late-type stars.
If instead $\rho(r)\sim r^{-7/4}$, $t_r$ continues to drop toward
Sgr A$^*$, as $t_r\sim r^{1/4}$.
Given the uncertainties, it is not clear that the relaxation time 
at the Galactic center is anywhere shorter than $10$ Gyr.

\subsection{Unrelaxed models}

\begin{figure}[!ht]
\plotfiddle{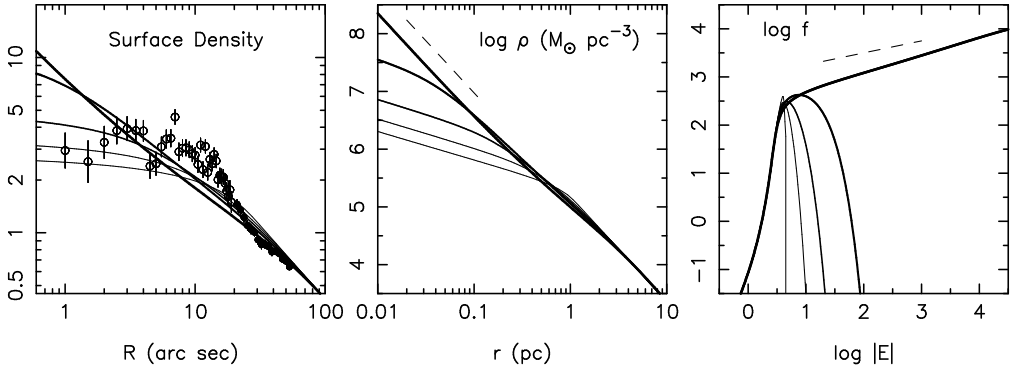}{1.75in}{270.}{55.}{55.}{-230}{250}
\caption{Evolution of the surface density $\Sigma(R)$,
configuration-space density $\rho(r)$,
and phase-space density $f(E)$ for a population of Solar-mass
stars around the MW SBH, assuming an intial core size of $1$ pc.
Increasing line thickness denotes increasing time,
$t=(0,0.2,0.5,1,2)\times 10^{10}$ yr.
Dashed lines are the asymptotic forms corresponding to a Bahcall-Wolf cusp,
i.e. $f\sim |E|^{1/4}$, $\rho\sim r^{-7/4}$.}
\end{figure}

Figure~2 shows the evolution of a single population of stars around 
the MW SBH, starting from a power-law density profile
(eq.~\ref{eq:rhoi}), with an initial core of radius $1$ pc.
The core ``fills in'' via gravitational encounters, on the
expected time scale of $t_r(r_\mathrm{infl})\approx 20$ Gyr.
By $5$ Gyr, the core has shrunk to a size of $\sim 0.5$ pc,
roughly the size of the core observed in the late-type stars
(Fig.~1).
Not until $\sim 20$ Gyr is a Bahcall-Wolf cusp fully established.

Because the density of the NSC beyond $r_\mathrm{infl}$ falls off as
$\rho\sim r^{-1.8}$ -- roughly the same, $r^{-7/4}$  dependence as
in a Bahcall-Wolf cusp -- the density in Figure~2 evolves in an
approximately ``self-similar'' way: the core shrinks while the
form of $\rho(r)$ outside the core always obeys
$\rho\sim r^{-1.8}$.
Initial cores in the range $1-1.5$ pc produce final cores,
after $5-10$ Gyr, that are consistent in size with the observed core.
The larger the initial core; the shorter the evolution time;
or the longer the relaxation time $t_r$, the larger the final core.

Why should there be a core in the first place?
Cores are ubiquitous features of luminous early-type galaxies;
core radii  are one to a few times  $r_\mathrm{infl}$,
consistent with formation via three-body ejection
of stars by the binary SBH that preceded the current, single SBH
\citep{Faber:97,MM:01}.
This model of core formation does not seem totally excluded 
for the Milky Way,
which might have experienced a major merger around the time
of formation of the thick disk, $10-12$ Gyr ago
\citep{Wyse:01}.
Furthermore the initial core size inferred above, $1-1.5$ pc,
is comparable to $r_\mathrm{infl}$.
But the core in the Milky Way is probably a different sort of
creature than the cores observed in luminous E-galaxies, since it sits at
the center of a nuclear star cluster.
Interestingly, the only other galaxy with a NSC that is
near enough for a parsec-scale core to be resolved -- the Local Group 
dwarf galaxy NGC 205 -- also contains a core, of radius $\sim 0.4$ pc
\citep{Valluri:05}.

Other ways of making a parsec-scale core include:

\begin{itemize}
\item {\sl Inspiral of intermediate-mass black holes} (IBHs).
A single IBH of mass $\sim 10^4M_\odot$,  spiralling in
against a pre-existing stellar density cusp,
creates a core of radius $\sim 0.05-0.1$ pc \citep{BGZ:06}.
Repeated inspiral events would create a larger core,
although the displaced mass increases at a less than linear
rate with the number of inspirals.
Nevertheless, some models postulate one such event
every $\sim 10^7$ yr \citep{Zwart:06}.

An inspiralling IBH was first proposed as a solution to
the other grand problem of the Galactic center, the origin
of the young stars (Hansen \& Milosavljevic 2003).
Subsequently, a very specific model was proposed for the
formation and runaway growth of an IBH in a dense, 
inspiralling star cluster \citep{GR:05}.
When the predictions of this particular model --
e.g., an extended tidal tail of young stars --
were not verified, the idea fell out of favor \citep{Paumard:09}.
But invoking an IBH still has much to recommend it.
For instance, an IBH is extremely efficient at randomizing the 
orbits of the S-stars,
and the transition radius between the S-stars and the clockwise
disk is roughly the expected  stalling radius for an IBH
 \citep{MGM:09}.

\item {\sl An enlarged loss cone.} 
Gravitational encounters drive a mass flux 
of $\sim M_\mathrm{SBH}/t_r(r_\mathrm{infl})$ into Sgr A$^*$.
The core that results from this diffusive loss process is very
small: its size is comparable to the radius of the capture sphere --
either the tidal disruption radius,
$r_t\approx 10^{-5}$ pc, or the Schwarzschild radius,
$r_\mathrm{Sch}\approx 10^{-6}$ pc.
The reason the core is so small is that the depleted orbits are
continuously resupplied by diffusion from
orbits of larger angular momentum and energy.
If there were some way to transfer a mass in stars of $\sim M_\mathrm{SBH}$
into the SBH on a time scale $\ll t_r$ -- say, a crossing time --
the resulting core would be much larger.
This could happen if the NSC were appreciably triaxial,
even if only transiently, since many orbits near a SBH in a triaxial cluster 
are ``centrophilic,''
passing arbitrarily close to the SBH after a finite time
\citep{MP:04}.

\item {\sl Localized star formation.} The phase-space density $f(E)$ of
an isotropic nucleus containing a core is roughly a delta-function
in energy, $f\sim \delta(E-E_0)$, with $E_0$ the gravitational potential
at the core radius.
(This can be seen in the initial conditions plotted in Fig.~2,
right panel).
Roughly the same initial conditions are implied by formation
of stars in a narrow ring at a radius $r_0$, where $\Phi(r_0)=E_0$, 
if it is assumed that the stellar orbital eccentricities and orientations
are randomized soon after the stars form.
The two, young stellar disks have mean radii of $\sim 0.25$ pc
and the clockwise disk extends inward as far as $\sim 0.05$ pc
\citep{Bartko:10},
but it is not out of the question that the bulk of star formation
took place in disks with radii $\sim 0.5$ pc or greater.
\end{itemize}

\begin{figure}[!ht]
\plotfiddle{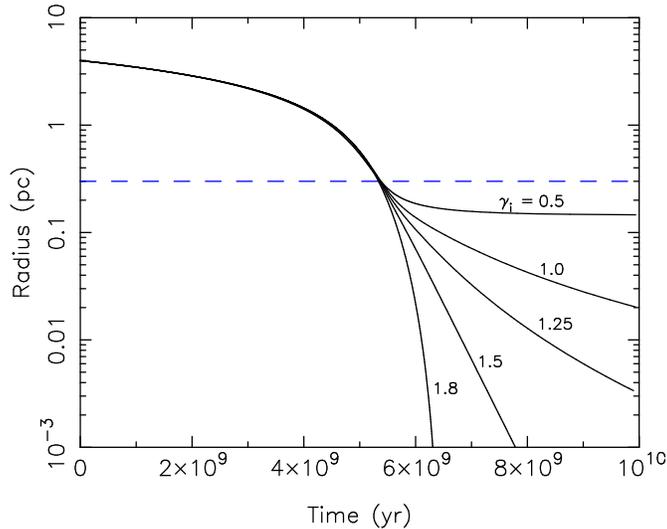}{2.40in}{0.}{45.}{45.}{-130}{-80}
\caption{Trajectories of $10 M_\odot$ BHs as they spiral in to the
Galactic center on circular orbits, starting from a radius of $4$ pc.
The assumed background density is a power-law, $\rho\propto r^{-1.8}$
at large radii, with an inner core (dashed line), and with
$\rho\sim r^{-\gamma}$ inside $r_c$.
}
\end{figure}

Even if the distribution of late-type giants is unrelaxed,
it is not necessarily the case that the stellar BHs also have
a low central density, since they would have spiralled in relative
to the stars \citep{Morris:93}.
However the inspiral time is a strong function of the stellar
density, since the latter determines the dynamical friction force.
Figure~3 shows inspiral times for $10 M_\odot$ BHs in models
of the NSC with stellar density $\rho\sim r^{-\gamma}$ inside the core;
thus $\gamma\approx 1.8$ corresponds to an unbroken power-law.
For $\gamma\lap 1$, inspiral slows dramatically at a radius of
$\sim r_c/2$; indeed, Chandrasekhar's formula implies that 
the frictional force vanishes completely, at $r\lap r_c/2$, when 
$\gamma=0.5$, although this prediction needs to be checked
via careful $N$-body simulations.

``Massive perturbers'' -- giant molecular clouds, star clusters,
etc. -- are present in the NSC at $r\gg r_c$,
and could scatter stars into the central parsec, 
at a potentially much higher rate than two-body
relaxation between Solar-mass stars \citep{Perets:07}.
Almost all of the scattered stars would be on orbits that are
unbound to the SBH; the density profile of these stars
would be $n\sim r^{-1/2}$ and their density near the SBH would
be low.
However, field binary stars that are deflected by massive perturbers 
onto eccentric orbits can undergo a three-body exchange
interaction with the SBH, resulting in capture of one of the
stars onto a tight orbit around the SBH.
The resultant radial distribution of the bound stars
will reflect the uncertain 
semi-major axis distribution of the parent binary population.
The rate of captures depends also on 
the binary fraction and on the distribution of perturber masses,
both of which are poorly known.
But estimates of the capture rate are as high as
$\sim 10^{-4}\mathrm{yr}^{-1}$ \citep{Perets:07}.
The low observed density of late-type stars in the inner parsec
places a limit on the effectiveness of this mechanism unless
most of the captured stars are too faint to be observed.

\subsection{Relaxed models}
Most dynamical models of the Galactic center published
in the last two decades fall into this category.
The relevance of such models to the Milky Way is called 
into question by the apparent absence of a Bahcall-Wolf cusp in the
old stars.

The existence of a density cusp could be reconciled with the
observed core if there is a change in the luminosity
function at roughly the core radius, such that the fraction of
bright giants is much smaller inside the core than outside.
For instance, the $1-3 M_\odot$ stars that are believed to dominate
the number counts at magnitudes $m_K\approx 15$ might never
have formed.
This hypothesis is consistent with the apparently top-heavy mass
function inferred for the stars in the two young 
stellar disks \citep{Bartko:10},
with the low integrated X-ray flux from the Galactic center
\citep{NS:05}, and with some theoretical expectations about the
mode of star formation near a SBH \citep{Naya:07}.
However, as noted above, such an IMF, if active over the entire
lifetime of the NSC, would  result in a much higher mass-to-light
ratio than observed in the central parsec \citep{LBK:09}.

Another possibility is that the giant stars have been selectively
destroyed  by collisions with other members of the Galactic center
population \citep{Genzel:96,Alexander:99,BD:99}.
In a relaxed, multi-mass cusp, the densities of  the light and
heavy components (e.g. main sequence stars, stellar BHs) 
follow $n\sim r^{-3/2}$ and $n\sim r^{-2}$ respectively; the 
BHs are predicted to dominate the mass density inside a
radius $\sim 0.01-0.1$ pc from Sgr A$^*$ \citep{HA:06,Freitag:06}.
In such a dense cusp, the probability that a given star will suffer 
a physical collision with another star, or stellar remnant, over its 
lifetime is very high inside $\sim 0.1$ pc \citep{Freitag:08}.
The observational consequences of such a collision are less clear.
Simulations suggest that in order to avoid evolving onto the red-giant 
branch,
a $1-3M_\odot$ star must lose more than $90\%$ of its mass
\citep{Dale:09}.
Even assuming a ``super-relaxed'' density cusp, in which the 
density of stellar BHs was arbitrarily increased to four times
its value in the relaxed models, Dale et al. (2009) found 
the rate of such collisions to be far too small to
explain the observed giant depletion.

\begin{figure}[!ht]
\plotfiddle{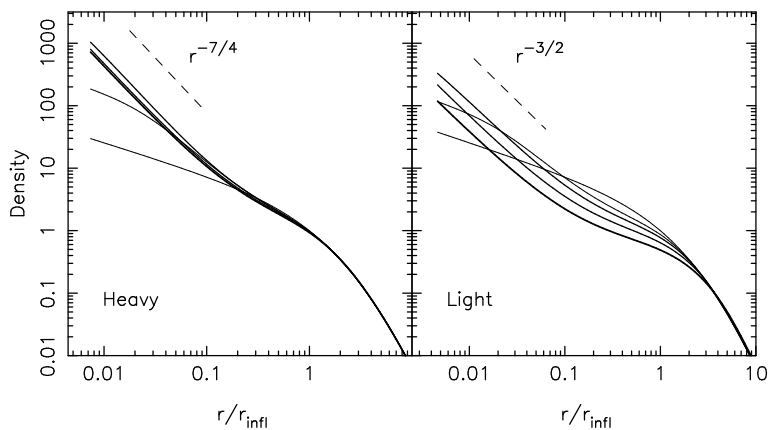}{2.15in}{-90.}{45.}{45.}{-160}{225}
\caption{Joint evolution of the density of stellar BHs (left) and low-mass
stars (right) around a SBH, under the assumption that the BHs
dominate the total density from the start.
Curves show densities at times $(0,0.25,0.5,1,2)$ in
units of the initial relaxation time at the influence radius.
The stars are scattered by the BHs into a $\rho\sim r^{-3/2}$ cusp, 
which retains its form as the density normalization drops with time, 
due to a continuous transfer of heat from the BHs.}
\end{figure}

At this meeting, M. Davies presented even more extreme models,
in which stars were assumed to form continuously from a flat IMF,
resulting in a core dominated by BHs throughout the inner parsec.
If the mass in BHs in the inner parsec is increased 
to several million Solar masses, 
the collision rate becomes high enough to
reproduce the observed giant depletion.
These models would appear to severely violate the 
proper-motion constraint on the total mass and the mass-to-light
ratio in the core \citep{SME:09}.

The collisional destruction model is nevertheless appealing,
and some way might still be found to make it work.
For instance, Dale et al. (2009) only considered collisions involving giants
approximately halfway up the giant branch;
for $2-3M_\odot$ stars, the red giant phase is so
short that collisions are more likely to occur before the
giant branch, in spite of the star's much smaller size
(J.  Lombardi, private communication).
Since the collisional probability is a strong function of
main-sequence mass,
precise spectral typing of the observed giants might provide
circumstantial evidence for such a model \citep{Do:09}.

It is sometimes argued \citep[e.g.][]{LBK:09}
that a high enough number of
stellar BHs could create a core, by sinking
to the center and displacing the less-massive stars.
Massive objects that dominate the local density
have two effects on the distribution of the
less-massive objects \citep[e.g.][]{MHB:07}.
There is a transfer of heat from the ``heavies'' (BHs) 
to the ``lights'' (stars), 
with a characteristic time given approximately by equation~(\ref{eq:tr}),
if $\tilde m$ is replaced by $m_\mathrm{BH}$.
In addition, the stars are scattered by the BHs, and
driven, on roughly the same time scale, 
to an approximately uniform population of phase
space.
A constant $f$ in the $1/r$ potential of a SMBH implies
a configuration-space density $n\sim r^{-3/2}$; thus the stars
would exhibit a
$r^{-3/2}$ cusp at $r\lap r_\mathrm{infl}$, 
the amplitude of which would gradually decay as the BHs continue
to heat the stars (Fig.~4).
In order to make a bona-fide core, the heavies must expel
the lights in a time $\ll t_r$; this happens, for instance,
when a second SBH spirals in.

\section{Conclusions}
To the long-standing ``paradox of youth'' 
at the Galactic center, 
we can now add a ``conundrum of old age'' arising from the
puzzling distribution of the late-type stars.
While the most recent data do not compel an interpretation of the
Galactic center as an unrelaxed system, they are broadly
consistent with such a model.
Even if the distribution of old stars is unrelaxed,
there might still be a ``dark cusp'' of stellar remnants
near Sgr A$^*$.
A key theoretical question is the efficiency of physical collisions at
keeping $1-3M_\odot$ stars from reaching the red giant branch.

\acknowledgements My understanding of the Galactic center is
based less on reading the relevant literature than on 
discussions with colleagues, particularly Tal Alexander, Don Figer and
Rainer Sch\"odel.
A. Graham, P. Kroupa, J. Lombardi and R. Sch\"odel 
read an early version of this report and provided helpful 
comments and corrections.
Some of the work described here was supported by grants AST-0807910 (NSF) 
and NNX07AH15G (NASA).



\begin{thebibliography}{}

\bibitem[Alexander(1999)]{Alexander:99} 
Alexander, T.\ 1999, \apj, 527, 835 

\bibitem[Bahcall \& Wolf(1976)]{BW:76} 
Bahcall, J.~N., \& Wolf, R.~A.\ 1976, \apj, 209, 214 

\bibitem[Bailey \& Davies(1999)]{BD:99} 
Bailey, V.~C., \& Davies, M.~B.\ 1999, \mnras, 308, 257 

\bibitem[Bartko et al.(2009)]{Bartko:09} 
Bartko, H., et al.\ 2009, \apj, 697, 1741 

\bibitem[Bartko et al.(2010)]{Bartko:10}
Bartko, H., et al. \ 2010, \apj, 708, 834 

\bibitem[Bastian et al.(2010)]{Bastian:10} 
Bastian, N., Covey, K.~R., \& Meyer, M.~R.\ 2010, arXiv:1001.2965 

\bibitem[Baumgardt et al.(2006)]{BGZ:06} 
Baumgardt, H., Gualandris, A., \& Portegies Zwart, S.\ 2006, \mnras, 372, 174 

\bibitem[B{\"o}ker(2008)]{Boeker:08} 
B{\"o}ker, T.\ 2008, 
Journal of Physics Conference Series, 131, 012043 

\bibitem[Buchholz et al.(2009)]{BSE:09} 
Buchholz, R.~M., Sch{\"o}del, R., \& Eckart, A.\ 2009, \aap, 499, 483 

\bibitem[Dale et al.(2009)]{Dale:09} 
Dale, J.~E., Davies, M.~B., Church, R.~P., \& Freitag, M.\ 2009, \mnras, 393, 1016 

\bibitem[Do et al.(2009)]{Do:09} 
Do, T., Ghez, A.~M., Morris, 
M.~R., Lu, J.~R., Matthews, K., Yelda, S., 
\& Larkin, J.\ 2009, \apj, 703, 1323 

\bibitem[Faber et al.(1997)]{Faber:97} 
Faber, S.~M., et al.\  1997, \aj, 114, 1771 

\bibitem[Figer(2004a)]{Figer:04a} 
Figer, D.~F.\ 2004a, 
The Formation and Evolution of Massive Young Star Clusters, 322, 49 
%
\bibitem[Figer et al.(2004b)]{Figer:04b} Figer, D.~F., Rich, 
R.~M., Kim, S.~S., Morris, M., \& Serabyn, E.\ 2004b, \apj, 601, 319 

\bibitem[Freitag et al.(2006)]{Freitag:06} 
Freitag, M., Amaro-Seoane, P., \& Kalogera, V.\ 2006, \apj, 649, 91 

\bibitem[Freitag et al.(2008)]{Freitag:08} 
Freitag, M., Dale, J.~E., Church, R.~P., \& Davies, M.~B.\ 2008, 
IAU Symposium, 245, 211 

\bibitem[Genzel et al.(1996)]{Genzel:96} 
Genzel, R., Thatte, N., Krabbe, A., Kroker, H., \& Tacconi-Garman, L.~E.\ 1996, \apj, 472, 153 

\bibitem[Genzel et al.(2003)]{Genzel:03} 
Genzel, R., et al.\ 2003, \apj, 594, 812 

\bibitem[Ghez et al.(1998)]{Ghez:98} Ghez, A.~M., Klein, B.~L., 
Morris, M., \& Becklin, E.~E.\ 1998, \apj, 509, 678 

\bibitem[Ghez et al.(2008)]{Ghez:08} 
Ghez, A.~M., et al.\ 2008, 
\apj, 689, 1044 

\bibitem[Gillessen et al.(2009)]{Gillessen:09} 
Gillessen, S., Eisenhauer, F., Trippe, S., Alexander, T., Genzel, R., Martins, F., 
\& Ott, T.\ 2009, \apj, 692, 1075 

\bibitem[G{\"u}rkan \& Rasio(2005)]{GR:05} 
G{\"u}rkan, M.~A., \& Rasio, F.~A.\ 2005, \apj, 628, 236 

\bibitem[Hansen \& Milosavljevi{\'c}(2003)]{HM:03} 
Hansen, B.~M.~S., \& Milosavljevi{\'c}, M.\ 2003, \apjl, 593, L77 

\bibitem[Hopman \& Alexander(2006)]{HA:06} 
Hopman, C., \& Alexander, T.\ 2006, \apjl, 645, L133

\bibitem[Hughes(2003)]{Hughes:03} 
Hughes, S.~A.\ 2003, Annals of Physics, 303, 142 

\bibitem[Levin et al.(2005)]{Levin:05} 
Levin, Y., Wu, A., \& Thommes, E.\ 2005, \apj, 635, 341 

\bibitem[L{\"o}ckmann et al.(2009)]{LBK:09} L{\"o}ckmann, U., 
Baumgardt, H., \& Kroupa, P.\ 2009, \mnras, in press

\bibitem[Lu et al.(2009)]{Lu:09} 
Lu, J.~R., Ghez, A.~M., Hornstein, S.~D., Morris, M.~R., Becklin, E.~E., 
\& Matthews, K.\ 2009, \apj, 690, 1

\bibitem[Merritt(2004)]{Merritt:04} 
Merritt, D.\ 2004, Physical Review Letters, 92, 201304 

\bibitem[Merritt(2009)]{Merritt:09} 
Merritt, D.\ 2009,  arXiv:0909.1318 

\bibitem[Merritt et al.(2009)]{MGM:09} 
Merritt, D., Gualandris, A., \& Mikkola, S.\ 2009, \apjl, 693, L35 

\bibitem[Merritt et al.(2007)]{MHB:07} 
Merritt, D., Harfst, S., \& Bertone, G.\ 2007, \prd, 75, 043517 

\bibitem[Merritt \& Poon(2004)]{MP:04} 
Merritt, D., \& Poon, M.~Y.\ 2004, \apj, 606, 788 

\bibitem[Merritt \& Szell(2006)]{MS:06} 
Merritt, D., \& Szell, A.\ 2006, \apj, 648, 890 

\bibitem[Mezger et al.(1999)]{Mezger:99} 
Mezger, P.~G., Zylka, R., Philipp, S., \& Launhardt, R.\ 1999, \aap, 348, 457 

\bibitem[Milosavljevi{\'c} \& Merritt(2001)]{MM:01} 
Milosavljevi{\'c}, M., \& Merritt, D.\ 2001, \apj, 563, 34 

\bibitem[Morris(1993)]{Morris:93} 
Morris, M.\ 1993, \apj, 408, 496 

\bibitem[Nayakshin et al.(2007)]{Naya:07} 
Nayakshin, S., Cuadra, J., \& Springel, V.\ 2007, \mnras, 379, 21 

\bibitem[Nayakshin \& Sunyaev(2005)]{NS:05} 
Nayakshin, S., \& Sunyaev, R.\ 2005, \mnras, 364, L23 

\bibitem[Oh et al.(2009)]{Oh:09} 
Oh, S., Kim, S.~S., \& Figer, D.~F.\ 2009, 
Journal of Korean Astronomical Society, 42, 17 

\bibitem[Paumard et al.(2006)]{Paumard:06} 
Paumard, T., et al.\ 2006, \apj, 643, 1011 

\bibitem[Paumard(2009)]{Paumard:09} 
Paumard, T.\ 2009, Journal of Physics Conference Series, 131, 012009 

\bibitem[Perets et al.(2007)]{Perets:07} 
Perets, H.~B., Hopman, C., \& Alexander, T.\ 2007, \apj, 656, 709 

\bibitem[Philipp et al.(1999)]{Philipp:99} 
Philipp, S., Zylka, R., Mezger, P.~G., Duschl, W.~J., Herbst, T., \& Tuffs, R.~J.\ 1999, \aap, 348, 768 

\bibitem[Portegies Zwart et al.(2006)]{Zwart:06} 
Portegies Zwart, S.~F., Baumgardt, H., McMillan, S.~L.~W., 
Makino, J., Hut, P., \& Ebisuzaki, T.\ 2006, \apj, 641, 319 

\bibitem[Preto et al.(2004)]{Preto:04} 
Preto, M., Merritt, D., \& Spurzem, R.\ 2004, \apjl, 613, L109 

\bibitem[Sch{\"o}del et al.(2007)]{Schoedel:07}
Sch{\"o}del, R., et al.\ 2007, \aap, 469, 125 

\bibitem[Sch{\"o}del et al.(2008)]{SME:08} 
Sch{\"o}del, R., Merritt, D., \& Eckart, A.\ 2008, 
Journal of Physics Conference Series, 131, 012044 

\bibitem[Sch{\"o}del et al.(2009)]{SME:09} 
Sch{\"o}del, R., Merritt, D., \& Eckart, A.\ 2009, 
\aap, 502, 91 

\bibitem[Valluri et al.(2005)]{Valluri:05} 
Valluri, M., Ferrarese, L., Merritt, D., \& Joseph, C.~L.\ 2005, 
\apj, 628, 137 

\bibitem[Wyse(2001)]{Wyse:01} 
Wyse, R.~F.~G.\ 2001, in {\it Galaxy Disks and Disk Galaxies}, 
Astronomical Society of the Pacific Conference Series,
vol. 230, ed. J.~G.~Funes \& E.~M.~Corsini,
71-80.

\end{thebibliography}
\end{document}